\documentclass[twocolumn, aps, prd, showpacs, nofootinbib]{revtex4-1}
\usepackage{graphics,graphicx}
\usepackage{amsmath,amssymb,amsfonts}

\begin{document}

\title{Compact stars in the braneworld: a new branch of stellar configurations with arbitrarily large mass}

\author{Germ\'an Lugones}\email{german.lugones@ufabc.edu.br}
\affiliation{Centro de Ci\^encias Naturais e Humanas, Universidade Federal do ABC, Avenida dos Estados 5001, 09210-580 Santo Andr\'e, S\~ao Paulo, Brazil.}

\author{Jos\'e D. V. Arba\~nil}\email{jose.arbanil@upn.pe}
\affiliation{Departamento de F\'{\i}sica, Instituto Tecnol\'ogico de Aeron\'autica, Centro T\'ecnico Aeroespacial, 12228-900 S\~ao  Jos\'e  dos Campos, S\~ao Paulo, Brazil.}
\affiliation{Departamento de Ciencias, Universidad Privada del Norte, Av. Alfredo Mendiola 6062 Urb. Los Olivos, Lima, Lima, Per\'u.}


\begin{abstract}

We study the properties of compact stars in the Randall-Sundrum II type braneworld model. To this end,  we solve the braneworld generalization of the stellar structure equations for a static fluid distribution with spherical symmetry considering that the spacetime outside the star is described by a Schwarzschild metric. 
First, the stellar structure equations are integrated employing the so called causal limit equation of state (EOS), which  is constructed  using a well established EOS at densities below a fiducial density, and the causal EOS $P= \rho$ above it. 
It is a standard procedure in general relativistic stellar structure calculations to use such EOS for obtaining a limit in the mass radius diagram, known as causal limit, above which no stellar configurations are possible if the EOS fulfills that the sound velocity is smaller than the speed of light. We find that the equilibrium solutions in the braneworld model can violate the general relativistic causal limit  and, for sufficiently large mass they approach asymptotically to the Schwarzschild limit $M = 2 R$.
Then, we investigate the properties of hadronic and strange quark stars using two typical  EOSs: a nonlinear relativistic mean-field model for hadronic matter and  the MIT bag model for quark matter.  For masses below $\sim 1.5 - 2 M_{\odot}$,  the mass versus radius curves show the typical behavior found within the frame of General Relativity. However, we also find a new branch of stellar configurations that can violate the general relativistic causal limit and that in principle may have an arbitrarily large mass.  The stars belonging to this new branch are supported against collapse by the nonlocal effects of the bulk on the brane. 
We also show that these stars are always stable under small radial perturbations.
{These results support the idea that traces of extra-dimensions might be found in astrophysics, specifically through the analysis of masses and radii of compact objects.}

\end{abstract}


\maketitle

\section{Introduction}

Braneworld models represent the universe as a three-dimensional brane where elementary particles live embedded in a higher-dimensional spacetime called the bulk, only accessed by gravity \cite{maartens2010}.
Within this  framework, astrophysical and cosmological models can be constructed where the gravitational effect of extra-dimensions can be assessed. Two well known examples of braneworld models are Randall-Sundrum (RS) \cite{Randall1999}  and Dvali-Gabadadze-Porrati (DGP) \cite{dvali2000} models. In RS models, ultraviolet modifications to General Relativity are introduced. Significant deviations from Einstein's theory occur at very high energies, e.g. in the early universe, in gravitational collapse and in compact objects. DGP models present infrared modifications with respect to General Relativity.

In the present work we focus on the RS type II braneworld model \cite{Randall1999}, which  has attracted much attention because it includes nontrivial gravitational dynamics despite a simple construction. In the RS model, our universe is a brane embedded in one extra dimension (the bulk)  which  is a portion of a 5D anti-de Sitter spacetime (AdS$_5$);  i.e. the extra dimension is curved or warped rather than flat. Significant deviations from Einstein's theory occur at very high energies. At low energies, gravity  has  an exponentially suppressed tail into the extra dimension due to a negative bulk cosmological constant, $\Lambda_5 = - 6/  \ell^2$ where $\ell$ is the curvature radius of AdS$_5$. The brane gravitates with self-gravity in the form of a brane tension $\lambda$,
where $\lambda = 3 M_p^2 / (4 \pi \ell^2)$ and $M_p^2 = M_5^3 \ell$.  On the brane, the negative $\Lambda_5$ is counterbalanced by the positive brane tension $\lambda$.

The Einstein's field equation  takes the conventional form but with an effective energy-momentum tensor $T^{\rm eff}_{\mu\nu}$, i.e., it reads:
\begin{equation}\label{eq_bm}
G_{\mu\nu}=8\pi G\,T^{\rm eff}_{\mu\nu},
\end{equation}
where $G_{\mu\nu}$ is the usual Einstein field tensor, and we consider $c=1$.

The effective energy-momentum tensor has the form  \cite{Shiromizu2000}
\begin{equation}
\label{teff}
    T_{\mu \nu }^{\mathrm{eff}}=T_{\mu \nu } + \frac{6}{\lambda}\,S_{\mu \nu }-\frac{1}{8\pi G}\,\mathcal{E}_{\mu \nu }\,,
\end{equation}
where the first  term contains the standard energy momentum tensor; e.g. for a perfect fluid we have  $T_{\mu \nu } = \rho u_{\mu }u_{\nu }+ph_{\mu \nu }$, where  $p$ is the pressure of the fluid,  $\rho$ is its energy
density,  $u^{\mu }$ is the four-velocity and $h_{\mu \nu }=g_{\mu \nu }+u_{\mu }u_{\nu }$ is the projection orthogonal to $u^{\mu }$. The second  and third terms include modifications with respect to the standard 4D Einstein's field equation. The bulk correction includes a local term  and a nonlocal one (second and third terms respectively) \cite{maartens2010}. For a perfect fluid, the local contribution reads
\begin{eqnarray}
 S_{\mu \nu } = \frac{1}{12} \rho^{2} u_{\mu } u_{\nu } + \frac{1}{12}  \rho(\rho+2p) h_{\mu \nu }.
\end{eqnarray}
The nonlocal contribution for static spherical symmetry reads:
\begin{equation}
    \mathcal{E}_{\mu \nu }=-\frac{6}{8\pi G\lambda}\,\left[ \mathcal{U}u_{\mu }u_{\nu }+
    \mathcal{P} r_{\mu } r_{\nu }+ \frac{(\mathcal{U}-\mathcal{P})}{3}  h_{\mu \nu } \right],
\end{equation}
where $\mathcal{U}$ and $\mathcal{P}$ are respectively the nonlocal energy density and nonlocal pressure on the brane and $r_{\mu }$ is a unit radial vector. Notice that,  as $\lambda \rightarrow \infty$, the bulk corrections vanish and General Relativity is recovered.

Some consequences for compact star physics  have been explored within the above described braneworld model. 
In their pioneering work,  Germani and Maartens \cite{maartens2001} showed that  the Schwarzschild solution is no longer the unique asymptotically flat vacuum exterior. In general, the exterior carries an imprint of nonlocal bulk graviton stresses and  knowledge of the 5D Weyl tensor is needed as a minimum condition for uniqueness. They also found an exact uniform-density stellar solution on the brane, and showed that the existence of neutron stars leads to a constraint on the brane tension that is stronger than the big bang nucleosynthesis constraint, but weaker than the Newton-law experimental constraint.
After this work, some other studies of spherical static stars considering a Schwarzschild exterior metric in the braneworld model were done in order to determine how local and nonlocal corrections affect the stellar structure.  In Refs. \cite{ovalle2008,ovalle2009,ovalle2010}  the role of the local and nonlocal corrections is examined in nonuniform and uniform stars. In these works, to overcome all problems associated with the system of equations,  the time metric component is prescribed. This approach helps to find an exact solution of the  Einstein's equations on the brane and an analytic form of the Weyl curvature terms. 
More recently, some  properties of compact stars in the braneworld model were analyzed  using neutron star equations of state and assuming that  the Weyl terms obey the simplest relation $\mathcal{P} = w \mathcal{U}$ \cite{castro2014}. However we must notice that in this work the boundary condition for the nonlocal energy density $\mathcal{U}$ was set at the stellar center and not at the surface of the star as it should be.

In this work we investigate several aspects of compact stars within braneworld models. 
In Sec. \ref{Stellar_structure} we review the stellar structure equations on the brane, the boundary conditions and explain the shooting method used for the numerical integration of the equations. 
In Sec. \ref{Section_causal_limit} we describe the causal limit equation of state (EOS) employed in the literature to obtain the causal limit above which compact stars are not expected in General Relativity. Thereafter, we employ the causal limit EOS to obtain such limit in braneworld models, finding that the general relativistic limit can be violated.   
In Sec. \ref{Models_hadronic_quark_stars} we study the properties of hadronic and quark stars using typical EOSs for hadronic and quark matter and find striking features that differentiate their structure with respect to the general relativistic case. 
Since only stellar configurations  in stable equilibrium are acceptable from the astrophysical point of view, we analyze in Sec. \ref{Stellar_stability} the stability of the compact stars under small radial perturbations using a static method that allows to determine the precise number of unstable normal radial modes. Finally, in  Sec.~\ref{conclusions} we  summarize and discuss our results.

\section{Stellar structure on the brane}
\label{Stellar_structure}

\subsection{Stellar structure equations and boundary conditions}

Germani and Maartens \cite{maartens2001}  solved the Einstein's equations on the brane and derived the braneworld generalization of the stellar structure equations for a static fluid distribution with spherical symmetry
\begin{eqnarray}
\frac{dm}{dr}&=&4\pi r^2\rho_{\mathrm{eff}},\label{eq001}\\
\frac{dp}{dr}&=&-(\rho+p)\frac{d\phi}{dr},\label{eq002}\\
\frac{d\phi}{dr}&=&\frac{Gm+4\pi Gr^3\left(p_{\mathrm{eff}}+\frac{4\mathcal{P}}{(8\pi G)^2\lambda}\right)}{r(r-2Gm)},\label{eq003}\\
\frac{d\mathcal{U}}{dr}  &=&  - (4\mathcal{U}+2\mathcal{P} )\frac{d\phi}{dr}    -2(4 \pi G)^2 (\rho+p)
\frac{d \rho}{dr}      \nonumber \\
 && -2\frac{d\mathcal{P}}{dr}-\frac{6}{r}\mathcal{P}\,,\label{eq004}
\end{eqnarray}
where
\begin{eqnarray}
&&\rho^{\rm eff}=\rho\left(1+\frac{\rho}{2\lambda}\right)+\frac{6\,{\cal U}}{(8\pi G)^2\lambda},\\
&&p^{\rm eff}=p + \frac{\rho}{2 \lambda}(\rho + 2 p)+\frac{2\,{\cal U}}{(8\pi G)^2\lambda}.
\end{eqnarray}
To solve Eqs. (\ref{eq001})$-$(\ref{eq004}) we need an equation of state $\rho = \rho(p)$ and a relation of the form $\mathcal{P} = \mathcal{P}(\mathcal{U})$   relating the nonlocal components (``dark'' equation of state).

Two of the boundary conditions of the stellar structure equations on the brane are the same as for the standard General Relativistic equations. Specifically, at the center of the star ($r=0$) the enclosed mass is zero: 
\begin{equation}
m(r=0) = 0 ,
\end{equation}
and at the surface of the  object the pressure vanishes:
\begin{equation}
p(R) = 0 .
\label{boundary_pressure}
\end{equation}
The remaining boundary condition is determined by the Israel-Darmois matching  condition $\left[G_{\mu\nu}r^{\nu}\right]_{\Sigma}=0$  at the surface of the object $\Sigma$, where $[f]_{\Sigma}\equiv f(R^{+})-f(R^{-})$ (in the following we use the superscripts $-$ and $+$ to indicate quantities inside and outside the star respectively). By the brane field equation \eqref{eq_bm}, this implies $[T^{\rm eff}_{\mu\nu} r^{\nu}]_{\Sigma}=0$, leading to $\left[p^{\rm eff} + {4 {\cal P}} / ({(8\pi G)^2 \lambda}) \right]_{\Sigma}=0$.
Since at the surface of the object we have $p(R)=0$, we have:
\begin{equation}
(4\pi G)^2\rho^2(R)+{\cal U}^-(R) +2 {\cal P}^-(R)= {\cal U}^+(R) + 2 {\cal P}^+(R) ,
\label{boundary3a}
\end{equation}
which  holds for any static spherical star with vanishing pressure at the surface.

In BW models, the Schwarzschild solution is no longer the unique asymptotically flat vacuum exterior.
In general, the exterior carries an imprint of nonlocal bulk graviton stresses. Knowledge of the 5D Weyl tensor is needed as a minimum condition for uniqueness.
If there are no Weyl stresses in the interior  (${\cal U}^- = {\cal P} ^- = 0)$, and if the energy density is non-vanishing at the surface, $\rho(R)\neq 0$, then there must be Weyl stresses in the exterior, i.e. the exterior solution cannot be the Schwarzschild one \cite{maartens2001}. Equivalently, if we assume a Schwarzschild exterior solution (${\cal U}^+ = {\cal P}^+ = 0$) and the energy density is nonzero at the surface, then the interior solution must have  nonvanishing  nonlocal Weyl stresses.

{On the other hand, despite previous no-go results, 
brane-world compact objects with a Schwarzschild exterior are obtained in Ref. \cite{ovalle2015}.}
This result is arises at the price of having negative pressure inside a narrow shell at the star surface,  which effectively acts as a solid crust separating the inner fluid from the vacuum exterior \cite{ovalle2015}. Such crust has a negligible thickness, falling below any physically sensible length scale for astrophysical sources, and the discontinuities in $\mathcal{U}$ and $\mathcal{P}$ at $r=R$ are negligibly small. Therefore, the crust can be neglected in the calculation of the mass-radius diagram and  other global stellar properties. 

{However, we emphasize that in the present work, the physicality of the Schwarzschild exterior is not necessarily guaranteed.}
Nonetheless, in order to simplify the analysis and to facilitate the comparison with GR, we focus here on a class of models that satisfy the following properties:

\begin{enumerate}
\item we consider a Schwarzschild exterior solution ($\mathcal{U}^+ = \mathcal{P}^+ = 0$); 
\item we assume $\mathcal{P}^-ˆ' = 0$, which is consistent with the isotropy of the physical pressure in the star.
\end{enumerate}

As a consequence, the interior must have nonvanishing nonlocal Weyl stresses ($\mathcal{U}^- \neq 0$). 
Therefore, the boundary condition for $\mathcal{U}$  at $r=R$ simplifies to:
\begin{equation}
(4\pi G)^2\rho^2(R)+{\cal U}^{-}(R)=0.
\label{boundary3b}
\end{equation}

In summary, the full set of equations to be solved is: 

\begin{eqnarray}
\frac{dm}{dr} & = & 4\pi\rho^{\rm eff}{r}^2 ,  \label{mass-brane}  \\
\frac{dp}{dr} & = & -(p+\rho)\frac{\left[4\pi Gp^{\rm eff}r+\frac{mG}{r^2}\right]}{\left[1-\frac{2mG}{r}\right]},\label{tov-brane} \\
\frac{d{\cal U}^{-}}{dr} & = & \frac{4{\cal U}^{-}}{p+\rho}\frac{dp}{dr} -2(4\pi G)^2(\rho+p)\frac{d\rho}{dr},\label{U-brane2}
\end{eqnarray}
with the boundary conditions $m(r=0) = 0$, $p(R) = 0$ and $(4\pi G)^2\rho^2(R)+{\cal U}^{-}(R)=0$. An equation of state $\rho = \rho(p)$ must be supplied to close the system. In the  limit $\lambda\rightarrow\infty$, we have $\rho^{\rm eff} \rightarrow \rho$ and $p^{\rm eff} \rightarrow p$, and the General Relativistic stellar structure equations are recovered.

\subsection{Numerical integration of the structure equations}
%
For a given EOS of the form $\rho = \rho(p)$ and a given value of the brane tension $\lambda$, Eqs. (\ref{mass-brane})$-$(\ref{U-brane2}) can be integrated simultaneously with a Runge-Kutta method from the center towards the surface of the object. However, since the boundary condition  for  ${\cal U}^{-}(r)$ is given at the star's surface, a \textit{shooting method} is used in order to match Eq. \eqref{boundary3b}.

The integration of Eqs. \eqref{mass-brane}, \eqref{tov-brane} and \eqref{U-brane2} begins with the values
\begin{equation}
m(0) = 0  , \quad p(0) = p_c, \quad  {\cal U}^{-}(0) = {\cal U}^{-}_{c, trial} ,
\end{equation}
where $p_c$ is a given value for the central pressure, and ${\cal U}^{-}_{c, trial}$ is 
a trial value of ${\cal U}^{-}$ at $r=0$. The integration proceeds outwards until the pressure vanishes in order to verify Eq. \eqref{boundary_pressure}. However,  after such integration Eq. \eqref{boundary3b} is not necessarily fulfilled. Therefore,  the trial value of ${\cal U}_c^{-}$ is corrected through a Newton-Raphson iteration scheme in order to improve the matching of Eq. \eqref{boundary3b} in the next integration.  The integration from $r=0$ is repeated successively until Eq. \eqref{boundary3b} is satisfied  with the desired precision.
Once such precision is attained,  the point at which the pressure of the fluid vanishes determines the star's radius $R$ and the star's mass $M= m(R)$.

It is worth mentioning that for some simple EOSs, Eq. \eqref{U-brane2} can be integrated analytically.  In Appendix \ref{appendixA} we derive the explicit solution for a linear EOS of the form $\rho= p / c_s^2 + b$, where $c_s^2$ and $b$ are arbitrary constants.  This EOS is very useful because it includes as special cases the \textit{causal} EOS $\rho=p$,  the \textit{ultra-relativistic} EOS $\rho=3 p$, and the \textit{MIT bag model} EOS for massless quarks $\rho = 3 p + 4 B$, that we will use below.

\section{Upper bound on the maximum mass of compact stars in the braneworld model: the causal limit}
\label{Section_causal_limit}

\subsection{The causal limit EOS}
\label{causal_limit_EOS}
A complete knowledge of the equation of state of neutron star matter is still a challenge at present. The EOS can be reliably determined up to $\sim 2\rho_{sat}$, being $\rho_{sat} \approx 151 \, \mathrm{MeV / fm^3}$ the nuclear saturation density.  
However, for larger densities, the determination of a well-founded EOS strongly depends on the knowledge of strong interactions in a regime that cannot be reached experimentally. 
As a consequence, there is a large amount of high-density EOSs  in the literature that incorporate several aspects that may play a crucial role at the inner core of the star, such as three-body forces, bosonic condensates, hyperonic degrees of freedom and quark  matter \cite{haensel,glen}.

An important aspect of neutron stars within the frame of General Relativity, is that there exists a maximum gravitational mass above which there are no stable stellar configurations. The maximum mass exists no matter what the EOS, but its determination depends on a deep comprehension of the EOS up to several times $\rho_{sat}$. However,  using the so called \textit{causal limit EOS}, it is possible to circumvent the uncertainties related to the properties of high-density matter and obtain upper bounds to the maximum allowed  mass of a neutron star \cite{haensel,glen}. The causal limit EOS can be constructed by using a detailed EOS at density ranges where they can be safely regarded as accurate and imposing generic constraints at densities exceeding some fiducial density, e.g., subluminal sound velocity and thermodynamic stability (see e.g. \cite{rhoades1972,haensel,glen}).  

In this work, we adopt the  well established Baym, Pethick, and Sutherland (BPS) EOS  \cite{bps} at densities below a fiducial density $\rho_t$, and a causal equation of state (i.e. sound velocity $=$ speed of light) $p = \rho - a$  above $\rho_t$ \cite{rhoades1972,glen}. Since both EOSs  are matched at an energy density $\rho_t$ and a pressure $p_t$,  the constant $a$ in the high density EOS is given by $a = \rho_t - p_t$, where $\rho_t$ and $p_t$ also fulfill the BPS EOS (see Table \ref{table_fiducial}).

\subsection{The causal limit in General Relativity}
\label{limit_GR}

\begin{table}[tb] 
 \begin{tabular*}{\linewidth}{c @{\extracolsep{\fill}} cc}
\hline\hline
$\rho_t[\rm MeV/fm^3]$ & $p_t[\rm MeV/fm^3]$ & $a[\rm MeV/fm^3]$\\\hline
       $192.6$         & $2.103$             & $190.5$ \\
       $217.9$         & $2.675$             & $215.2$ \\
       $260.1$         & $3.809$             & $256.3$ \\
       $285.8$         & $4.613$             & $281.2$ \\
\hline\hline
\end{tabular*}
\caption{Values of the fiducial energy density $\rho_t$, fiducial pressure $p_t$, and  $a \equiv \rho_t-p_t$.   The  values of $\rho_t$ and $p_t$ were extracted from Table~V of Ref. \cite{bps}.}
\label{table_fiducial} 
\end{table}

\begin{figure}[tb]
\includegraphics[scale=0.30]{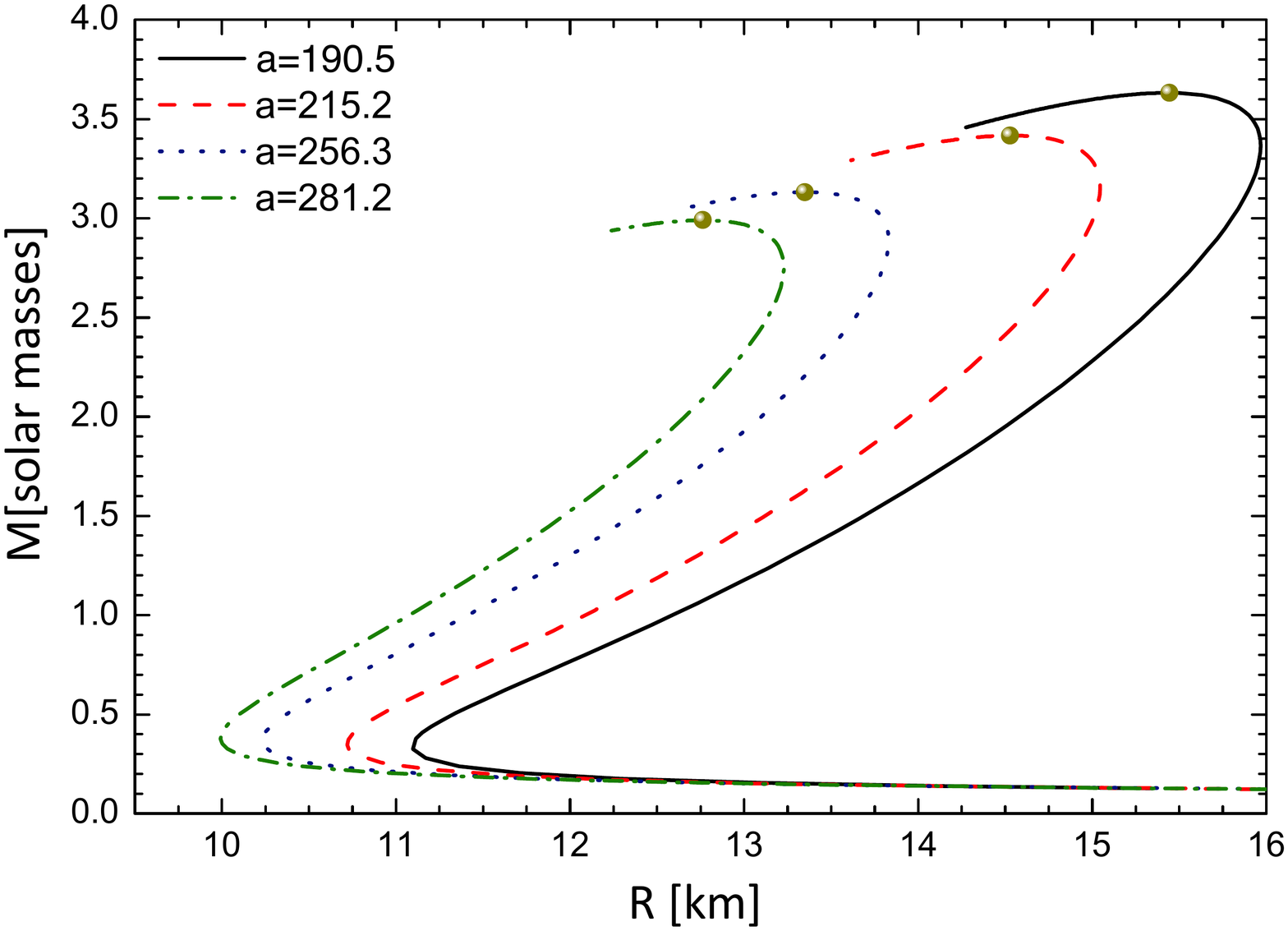}
\caption{Mass-radius relationship in General Relativity ($\lambda \rightarrow \infty$) for the causal limit EOS
matched continuously with the BPS EOS. Both EOSs  are matched at different fiducial densities that lead to different values of $a=\rho_f-p_f$. The dots over the curves indicate the maximum masses, which fall along the causal limit of Fig. \ref{LC}. }
\label{LC_GR}
\end{figure}

Using the causal limit EOS, it is possible to obtain a curve, know as \textit{causal limit}, that represents an upper bound in the mass-radius diagram for compact stars. The procedure to find the causal limit  within the frame of General Relativity has been explained in several textbooks (see e.g. \cite{glen,haensel}). For completeness, we present it here and in the next subsection we discuss it within the braneworld model.

For a given value of $a$, the stellar structure equations can be integrated, and a maximum stellar mass can be determined together with the corresponding stellar radius. For example, using $\rho_t=260.1\,{\rm MeV/fm^3}$ and $p_t=3.809\,{\rm MeV/fm^3}$), the sequence has a maximum mass object with:
\begin{equation}
M_{max}=3.131\,{\rm M_{\odot}} , \qquad R = 13.35 \,{\rm km}.
\end{equation}
Repeating the calculations for many different values of $a$, it can be shown that the maxima fall on a straight line given by  $M =  0.345 R$ (see Fig.~\ref{LC_GR}).

Therefore, the region excluded by causality in the $M-R$ diagram is given by \cite{glen,haensel,lattimer2004,lattimer2012}:
\begin{equation}
M \gtrsim  0.345 R  ,
\label{causal_limit_GR_2}
\end{equation}
or, equivalently:
\begin{eqnarray}
\left( \frac{M}{M_{\odot}} \right) \gtrsim 0.234 \left( \frac{R}{\mathrm{km} }  \right) .
\label{causal_limit_GR}
\end{eqnarray}
%

\subsection{The causal limit in the braneworld model}

\begin{figure}[tbh]
\centering
\includegraphics[scale=0.24]{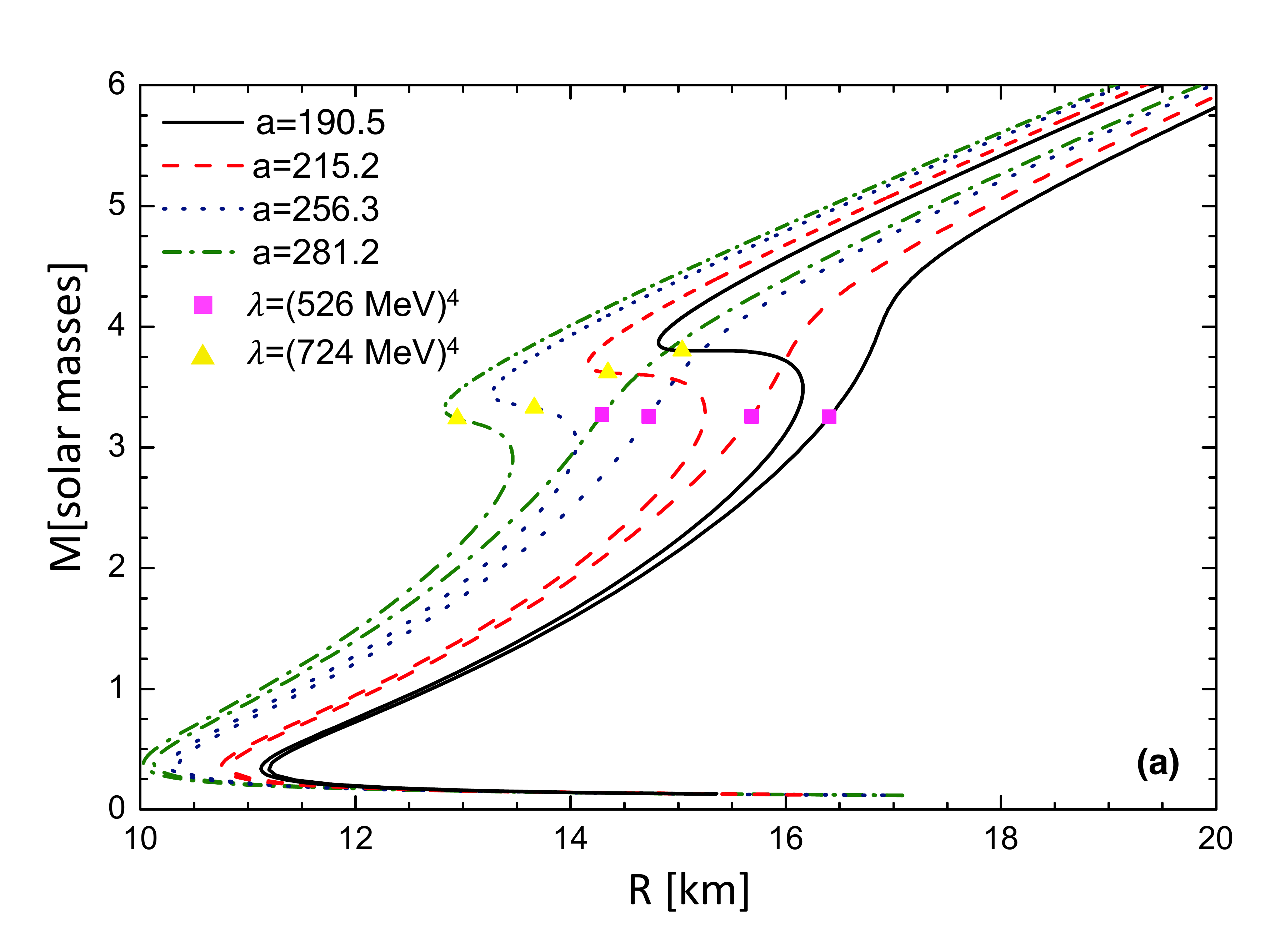}
\centering
\includegraphics[scale=0.24]{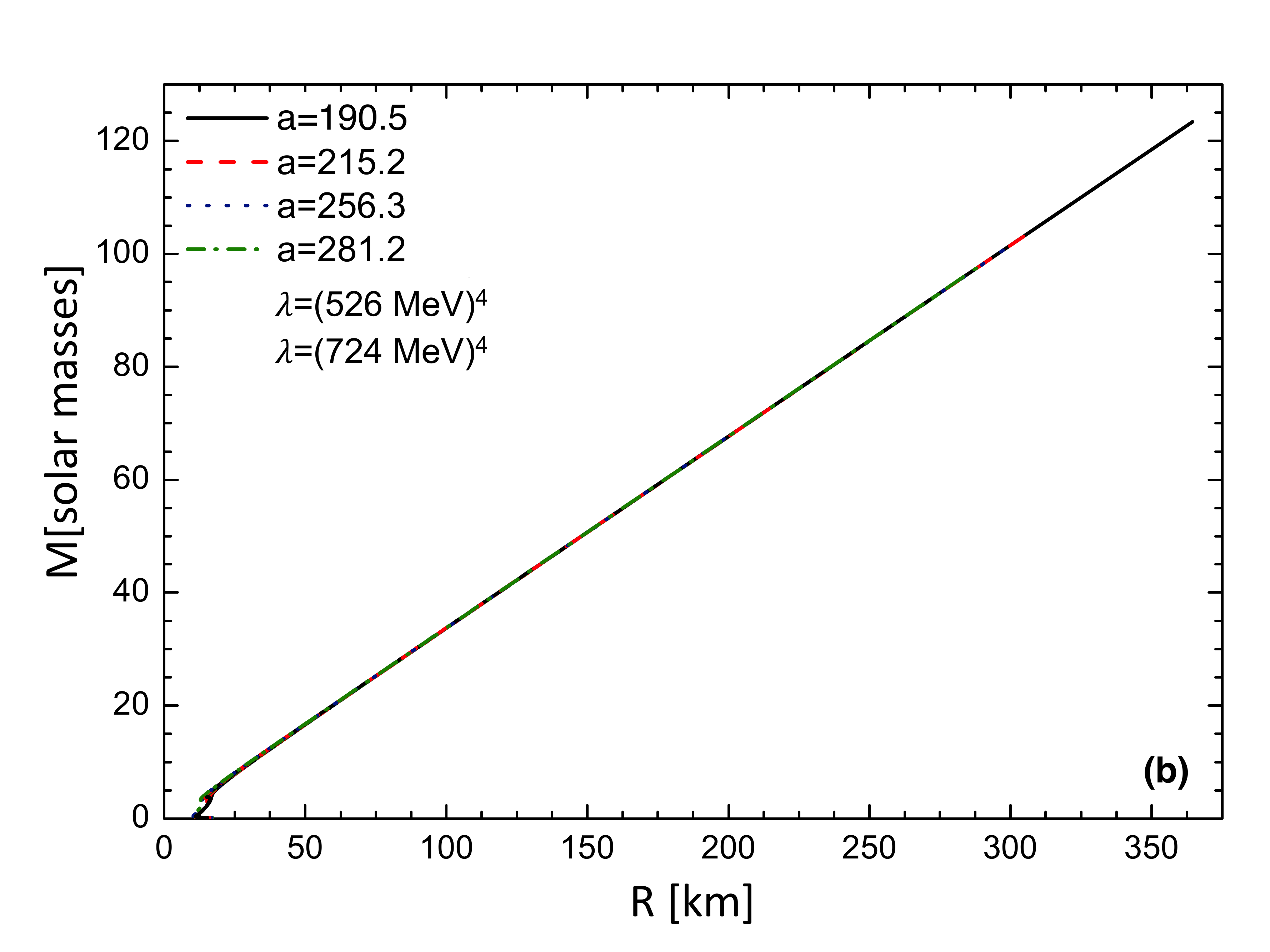}
\caption{(a) The mass-radius relationship  obtained using the causal limit EOS given  in Sec. \ref{causal_limit_EOS} for some values of $a$ (in ${\rm MeV/fm^3}$) and two different values of $\lambda$.
(b) Same as in (a) but for a larger range of $M$ and $R$. }
\label{LC_branes}
\end{figure}

\begin{figure}[tbh]
\centering
\includegraphics[scale=0.33]{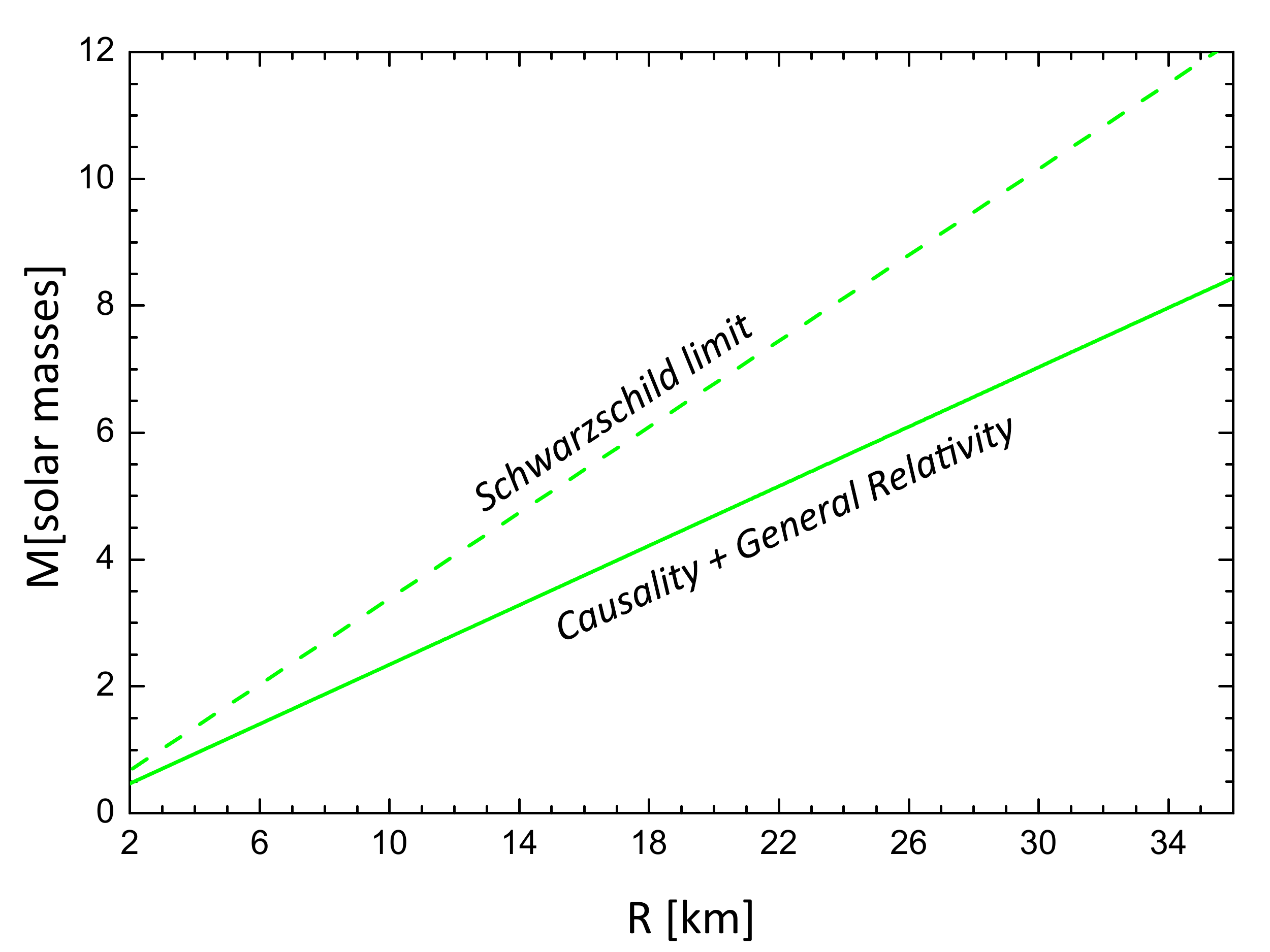}
\caption{The causal limit for General Relativity and the Schwarzschild limit $M = 2 R$. In the braneworld model, static stellar configurations fulfilling a causal EOS (sound velocity $<$ speed of light) can occupy the region between both straight lines.}
\label{LC}
\end{figure}

In order to determine the causal limit in the braneworld model, we integrate the stellar structure equations on the brane using the causal limit equation of state presented in Sec. \ref{causal_limit_EOS}. For small masses ($\lesssim 1.5 - 2 M_{\odot}$),  the curves show the typical behavior found within the frame of General Relativity. Specifically, very small mass stars have very large radii, and as the mass increases above a few tenths of solar masses the radii fall within a range of few kilometers around $\sim 10$ km. Nevertheless, for large mass objects,  local high-energy effects as well as nonlocal corrections lead to significant deviations with respect to General Relativity. At around $1.5 - 2 M_{\odot}$ the $M(R)$ curves bend anticlockwise as in the general relativistic case. However, instead of reaching a maximum mass as in General Relativity, the curves bend once more (clockwise) for larger masses and thereafter they increase roughly linearly (see Fig.~\ref{LC_branes}).  

A striking feature of this behavior  is that once the $M - R$ curves bend clockwise they may  fall above the causal limit obtained within General Relativity (c.f. Eqs. (\ref{causal_limit_GR}) $-$ (\ref{causal_limit_GR_2})). It can also be checked that  as the masses and radii increase, the curves tend asymptotically to the Schwarzschild limit $M = 2 R$. The asymptotic approach depends on the value of the brane tension $\lambda$: when $\lambda$ is small the curves go close to the line $M = 2 R$ at relatively small masses; but, for large $\lambda$  the approaching occurs at higher masses.

Since the $M - R$ curves for the causal EOS approach asymptotically the line $M = 2 R$, but do not go beyond it,   the Schwarzschild limit $M = 2 R$ is a good representation of the causal limit in the braneworld model ~\footnote{In a more general way,  this is also discussed  in \cite{Dvali2011}. They focus on  the field theoretical description of a generic theory of gravity flowing to Einstein General Relativity in IR and prove that, if ghost-free, in the weakly coupled regime such a theory can never become weaker than General Relativity.}. In other words, the equilibrium solutions found in the braneworld can violate the limit of causality for General Relativity (Eqs. (\ref{causal_limit_GR}) $-$ (\ref{causal_limit_GR_2})) and, for sufficiently large mass, can occupy the region between the straight lines shown in Fig.~\ref{LC}.

\section{Models for hadronic and strange quark stars}
\label{Models_hadronic_quark_stars}

In this section, we investigate the properties of hadronic and strange quark stars using typical models for the equations of state. As mentioned in Sec. \ref{causal_limit_EOS} there is a large amount of high density EOS that fulfill present experimental and astrophysical constrains. However, our purpose is not making an exhaustive survey of all the available EOSs, but rather to explore the qualitative properties of hadronic and strange quark stars using two models that have been extensively employed in the literature: a nonlinear relativistic mean-field model for hadronic matter and  the MIT bag model for quark matter.  In Sec. \ref{Equations_of_state} we briefly summarize the EOSs and in Sec. \ref{Structural_properties} we study the structural properties of the resulting compact objects.

\subsection{Equations of state}
\label{Equations_of_state}

For the hadronic phase we use a non-linear Walecka model \cite{Walecka,Walecka2,glendenning1991}
including the whole baryon octet, electrons and the corresponding antiparticles. The Lagrangian is given by
\begin{equation}
{\cal L}={\cal L}_{B}+{\cal L}_{M}+{\cal L}_{L}, \label{octetlag}
\end{equation}
where the indices $B$, $M$ and $L$ refer to baryons, mesons and leptons respectively. For the baryons we have
\begin{eqnarray}
{\cal L}_B &=&  \sum_B \bar \psi_B \bigg[\gamma^\mu\left
(i\partial_\mu - g_{\omega B} \ \omega_\mu- g_{\rho B} \ \vec \tau
\cdot \vec \rho_\mu \right) \nonumber \\
      &   & -(m_B-g_{\sigma B} \ \sigma)\bigg]\psi_B,
\end{eqnarray}
with $B$ extending over nucleons $n$, $p$ and the following hyperons $\Lambda$, $\Sigma^{+}$, $\Sigma^{0}$, $\Sigma^{-}$, $\Xi^{-}$, and $\Xi^{0}$. The contribution of the mesons $\sigma$, $\omega$ and $\rho$ is given by
\begin{eqnarray}
{\cal L}_{M} &=& \frac{1}{2} (\partial_{\mu} \sigma \ \!
\partial^{\mu}\sigma -m_\sigma^2 \ \! \sigma^2) - \frac{b}{3} \ \!
m_N\ \! (g_\sigma\sigma)^3 -\frac{c}{4} \ (g_\sigma \sigma)^4
\nonumber\\
& & -\frac{1}{4}\ \omega_{\mu\nu}\ \omega^{\mu\nu} +\frac{1}{2}\
m_\omega^2 \ \omega_{\mu}\ \omega^{\mu}       \nonumber\\
&  & -\frac{1}{4}\ \vec \rho_{\mu\nu} \cdot \vec \rho\ \! ^{\mu\nu}+
\frac{1}{2}\ m_\rho^2\  \vec \rho_\mu \cdot \vec \rho\ \! ^\mu,
\end{eqnarray}
and the coupling constants are
\begin{eqnarray}
g_{\sigma B}=x_{\sigma B}~ g_\sigma,~~g_{\omega B}=x_{\omega B}~
g_{\omega},~~g_{\rho B}=x_{\rho B}~ g_{\rho}.
\end{eqnarray}
Electrons are included as a free Fermi gas, ${\cal L}_{L}=\sum_l \bar \psi_l \left(i \rlap/\partial -
m_l\right)\psi_l$, in chemical equilibrium with all other particles.

The constants in the model are determined by the properties of nuclear matter and hyperon potential depths known from hypernuclear experiments. In the present work we use the GM1 parametrization for which we have $(g_{\sigma}/m_{\sigma})^{2} = 11.79$ fm$^{-2}$, $(g_{\omega}/m_{\omega})^{2} = 7.149$ fm$^{-2}$, $(g_{\rho}/m_{\rho})^{2} = 4.411$ fm$^{-2}$, $b = 0.002947$ and $c = 0.001070$  \cite{glendenning1991}. For the hyperon coupling constants we adopt $x_{\sigma i} = x_{\rho i} = 0.6$ and $x_{\omega i} = 0.653$  \cite{glendenning1991}.  At low densities we use the Baym, Pethick and Sutherland (BPS) model \cite{bps}.
For details on the explicit form of the equation of state derived from this Lagrangian the reader is referred to Refs. \cite{lugones_eos1,lugones_eos2} and references therein.


We describe quark matter through the MIT bag model. For simplicity we assume a zero strong coupling constant and consider massless quarks. If such effects were taken into account, the equation of state would be qualitatively the same but we would find nonanalytic expressions. In practice, only $u$, $d$, and $s$ quarks appear in quark matter because other quark flavors have masses much larger that the chemical potentials involved (roughly $300$ MeV). Since these quarks are assumed to be massless, leptons are not necessary to electrically neutralize the phase, and thus, they are not present in the system \cite{farhi1984}. In such a case, the equation of state adopts the simple form
\begin{equation}\label{eos_quark}
 \rho = 3 p + 4 B  ,
\end{equation}
where $B$ is the bag constant. Witten \cite{witten1984} conjectured that, at zero pressure and temperature, three flavor quark matter may have an energy per baryon smaller than ordinary nuclei. This would make strange quark matter the true ground state of strongly interacting matter and would lead to the existence of strange quark stars i.e. stellar objects completely  composed by strange quark matter \cite{Alcock1986}.  Within the MIT bag model for massless quarks and zero strong coupling constant, the Witten hypothesis is verified if the bag constant is in the range $57 \,{\rm MeV/fm^3} \lesssim B \lesssim 94 \,{\rm MeV/fm^3}$. In this paper we adopt $B = 60\,{\rm MeV/fm^3}$.

\subsection{Structural properties of hadronic and strange quark stars}
\label{Structural_properties}

In the following we present our results for hadronic and strange quark stars using the equations of state presented in the previous subsection.


\begin{figure}[tb]
\includegraphics[scale=0.24]{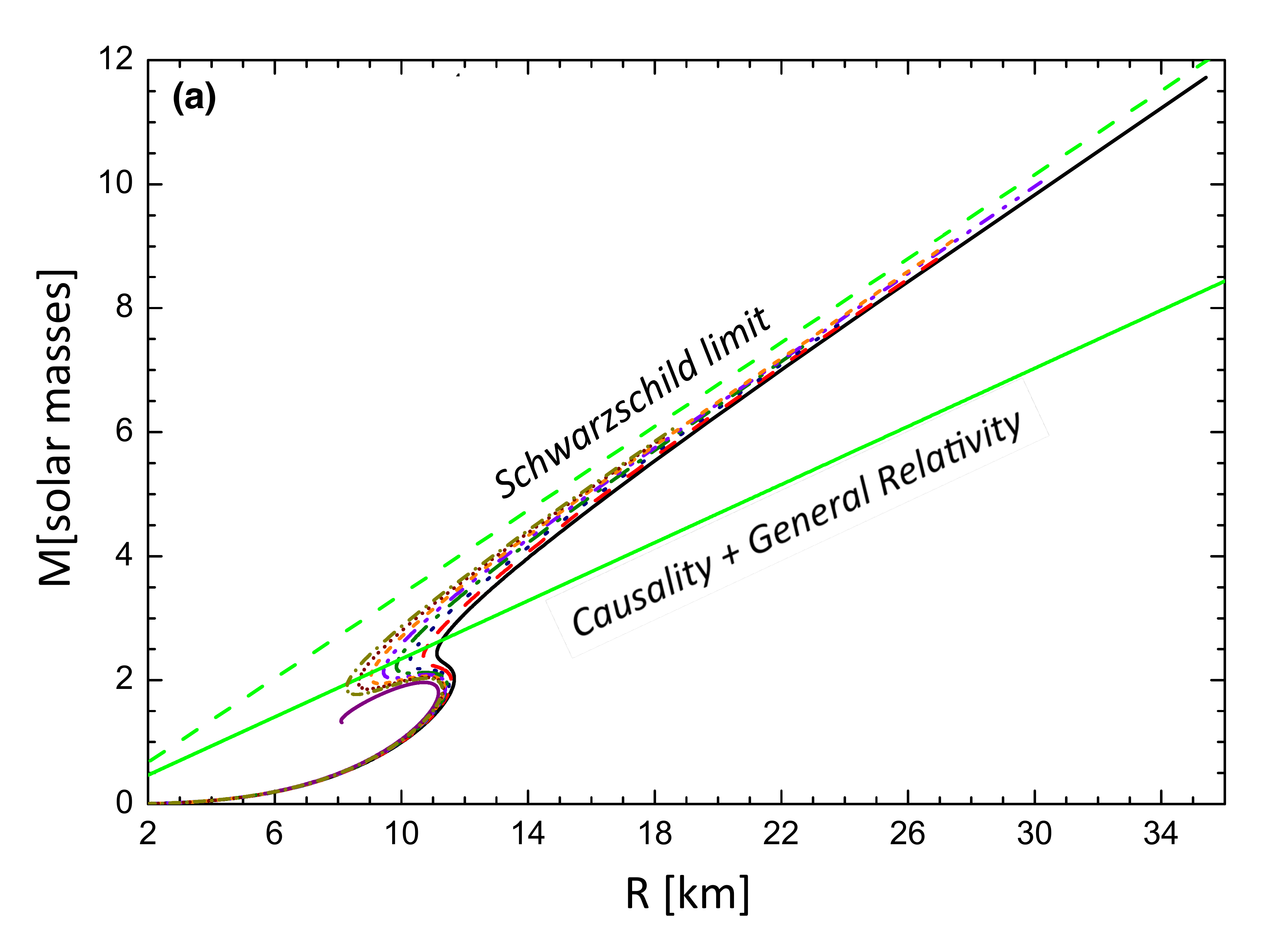}
\includegraphics[scale=0.24]{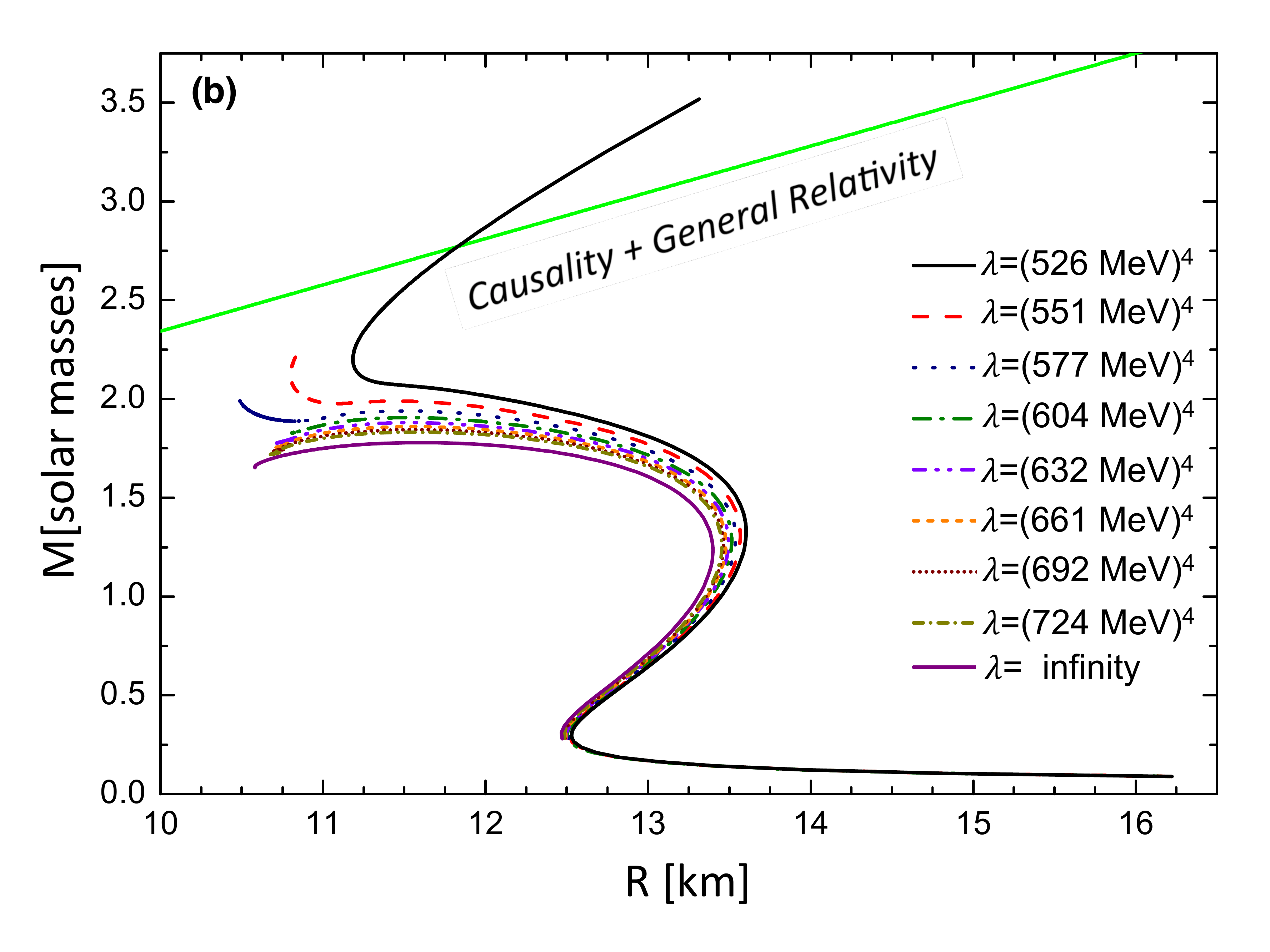}
\caption{Mass-radius relationship for (a) strange quark stars and (b) hadronic stars,  using  several values of the brane tension $\lambda$.
{These values of  $\lambda$ lead to $M_5 = (\tfrac{4}{3} \pi \lambda M_p^2)^{1/6} \sim 2000 \, \mathrm{TeV}$; i.e. larger than 10 TeV, in compatibility with LHC.}
We also show the general relativistic and braneworld model causal limits.}
\label{M_R}
\end{figure}

In Fig.~\ref{M_R} we show the mass-radius relationship for some values of the brane tension $\lambda$.  At the top panel we display the results for strange quark matter and at the bottom panel for hadronic matter. We also include the causal limit found before for General Relativity and the Schwarzschild limit $M = 2 R$.

For small masses ($\lesssim 1.5 - 2 M_{\odot}$),  the curves show the typical behavior found within the frame of General Relativity, i.e. very small mass hadronic stars have very large radii,  while strange stars follow roughly $M(R)  \propto R^3$. 

For large mass objects, braneworld effects lead to the deviations with respect to General Relativity that were explained in the case of the causal EOS of previous section.  At around $1.5 - 2 M_{\odot}$ the $M(R)$ curves for hadronic and quark stars  bend anticlockwise as in the general relativistic case. But then, the curves bend once more (clockwise) for larger masses and thereafter they increase roughly linearly and approach asymptotically to the Schwarzschild limit.  

In summary, the main characteristics of the mass-radius relationship already found for the causal EOS are confirmed for both the hadronic and the strange quark mater EOSs: 

\begin{itemize}
\item The $M(R)$ curves violate the general relativistic causal limit for large enough masses; instead, they can occupy the region between the  general relativistic causal limit and  the Schwarzschild limit.

\item Static stellar configurations do not have a maximum mass as in the  general relativistic case, i.e. objects of any mass are possible in principle.  

\end{itemize}

Notice that, differently from the causal EOS, we find now that in some cases there is a local maximum in the $M(R)$ curves at $M \sim 2 M_{\odot}$. Nevertheless, after bending clockwise, the behavior of all the $M(R)$ curves is qualitatively the same.


\begin{figure}[tb]
\centering
\includegraphics[angle=90,scale=0.3]  {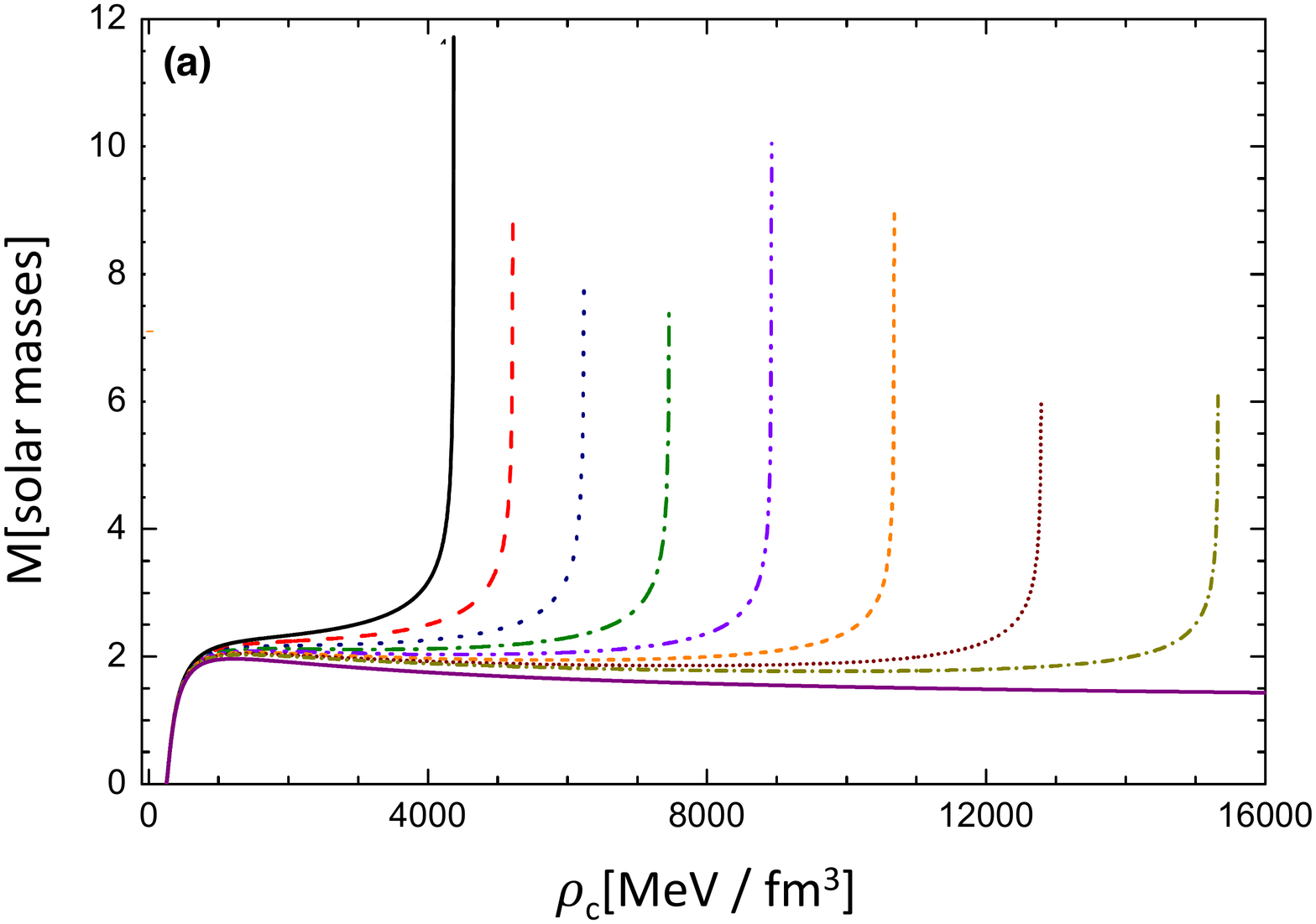}
\centering
\includegraphics[angle=90,scale=0.3]   {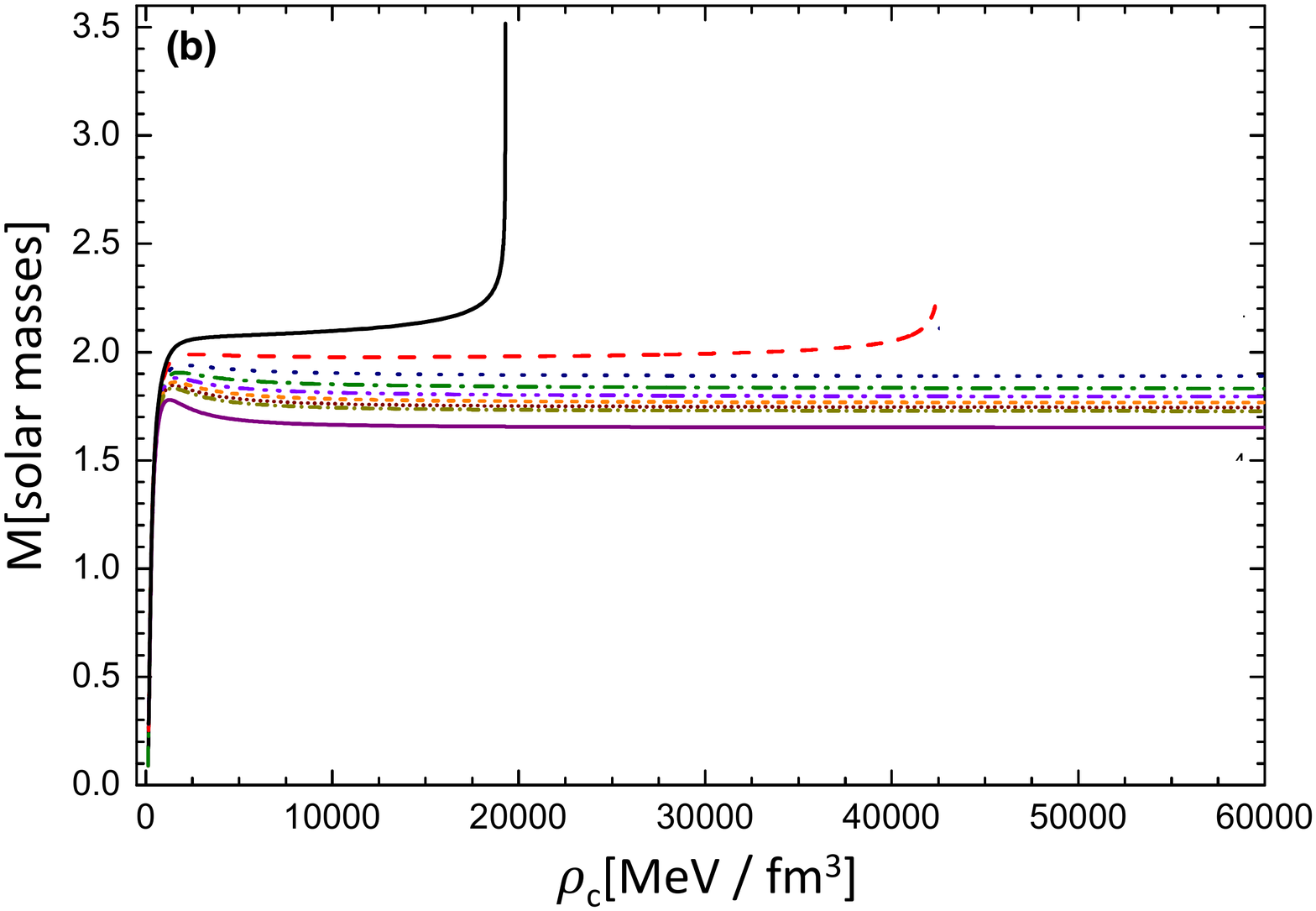}
\vspace*{-.3cm}
\caption{Mass of (a) strange stars and (b) hadronic stars versus the central energy density $\rho_c$ for different values of brane tension $\lambda$. The labels of the curves are the same as given in Fig. \ref{M_R}. }
\label{M_rhoc}
\end{figure}

Fig. \ref{M_rhoc} shows the dependence of the mass  with the central energy density  $\rho_c$ for some values of the brane tension $\lambda$. For a given value of $\rho_c$, the mass of a star is larger in the braneworld model  than in General Relativity  due to local and nonlocal extra-dimensional modifications to the structure equations on the brane. As expected, these corrections are small for low central energy densities but they become significant with increasing central energy density, specially for the smaller values of the brane tension $\lambda$. 
A remarkable feature of the $M(\rho_c)$ curves is that there is a value of $\rho_c$ for which the stellar mass diverges. 
This means that for large enough masses the nonlocal energy density ${\cal U}^{-}$ supports the star against gravitational collapse.
The maximum value of $\rho_c$ increases with the brane tension $\lambda$ as can be seen in Fig. \ref{M_rhoc}. In particular, as we approach to the general relativistic case ($\lambda \rightarrow \infty$) the maximum density is shifted to infinity.

\begin{figure}[tb]
\centering
\includegraphics[scale=0.24]{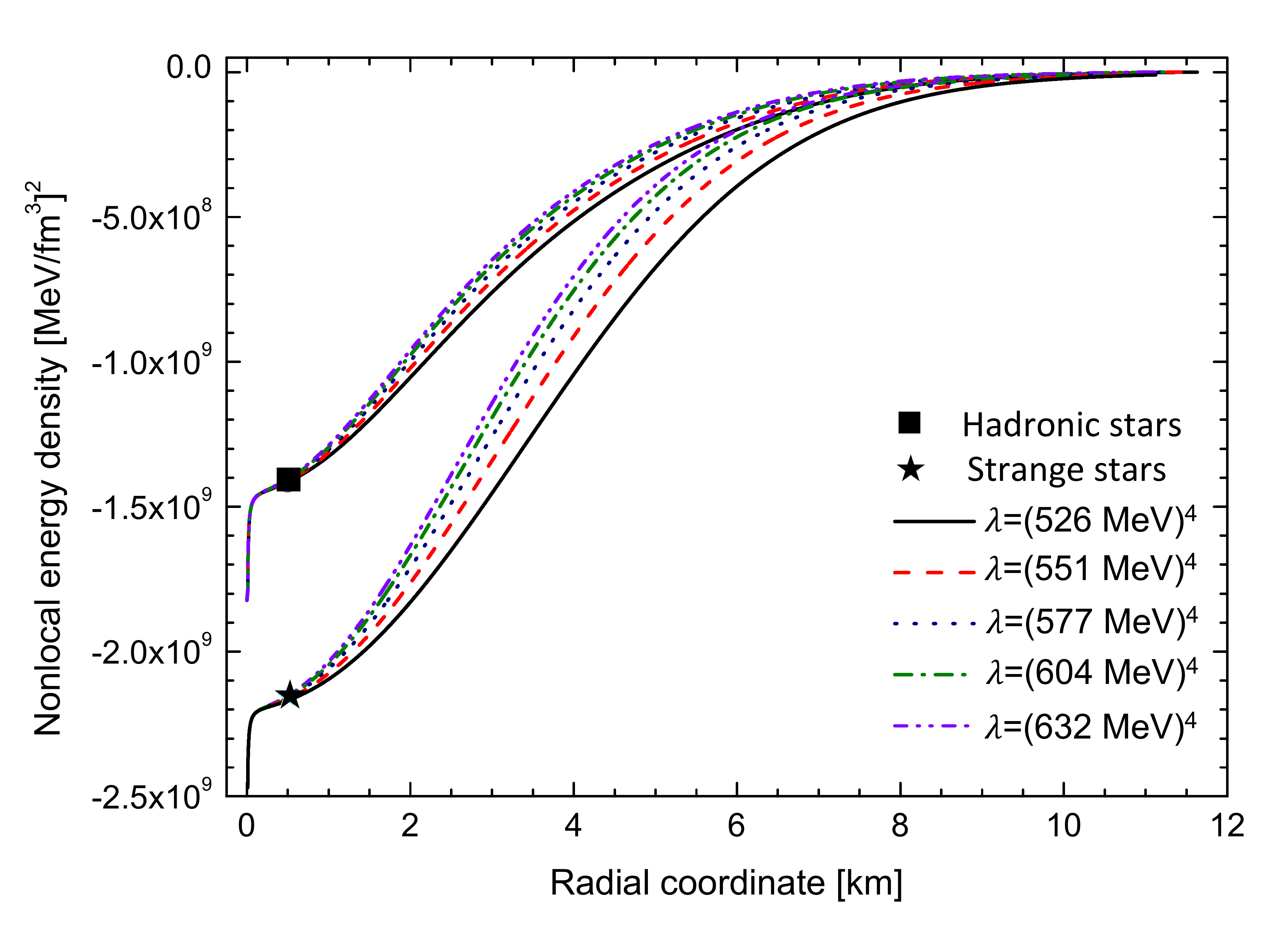}
\caption{Nonlocal energy density as a function of the radial coordinate for strange quark stars and hadronic stars. In all cases the central energy density is $\rho_c = 2500\,{\rm MeV/fm^3}$.}
\label{r_U}  
\end{figure}

In Fig. \ref{r_U} we show  the nonlocal energy density  ${\cal U}^{-}$  as a function of the radial coordinate $r$ for a central energy density  $\rho_c =2500\,{\rm MeV/fm^3}$ and five values of the brane tension $\lambda$. 
For both,  strange quark stars and hadronic stars,  the nonlocal energy density starts  at a large negative value at the center of the star and grows  monotonically towards the stellar surface. The more negative values of ${\cal U}^{-}$ are found for the lower values of $\lambda$.  A star with a more negative nonlocal energy density  admits more mass, because ${\cal U}^{-}$  acts as an effective negative pressure helping against the collapse.

\section{Stellar stability}
\label{Stellar_stability}

\begin{figure*}[tb]
\centering
\includegraphics[scale=0.24]  {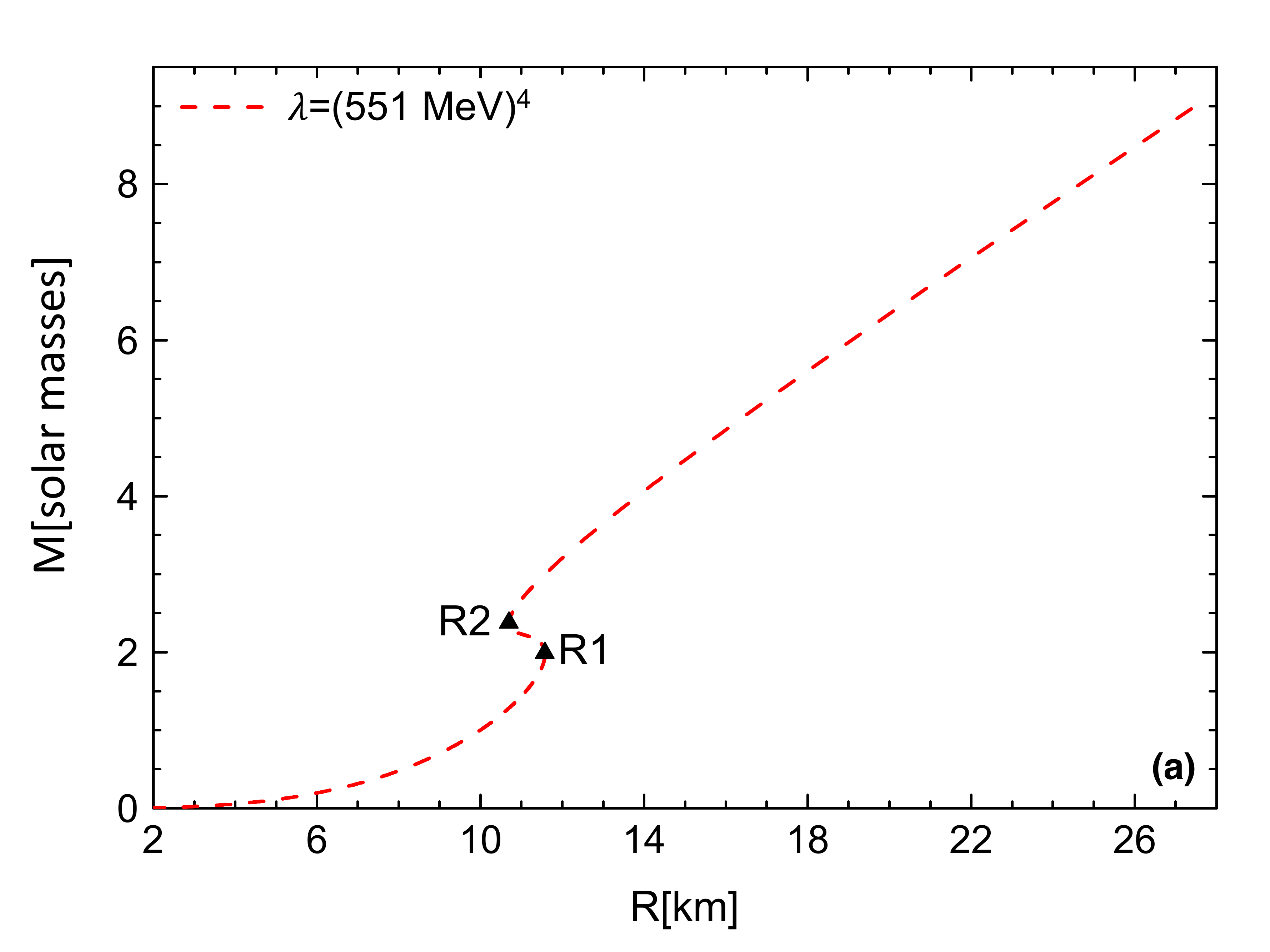}
\includegraphics[scale=0.24]  {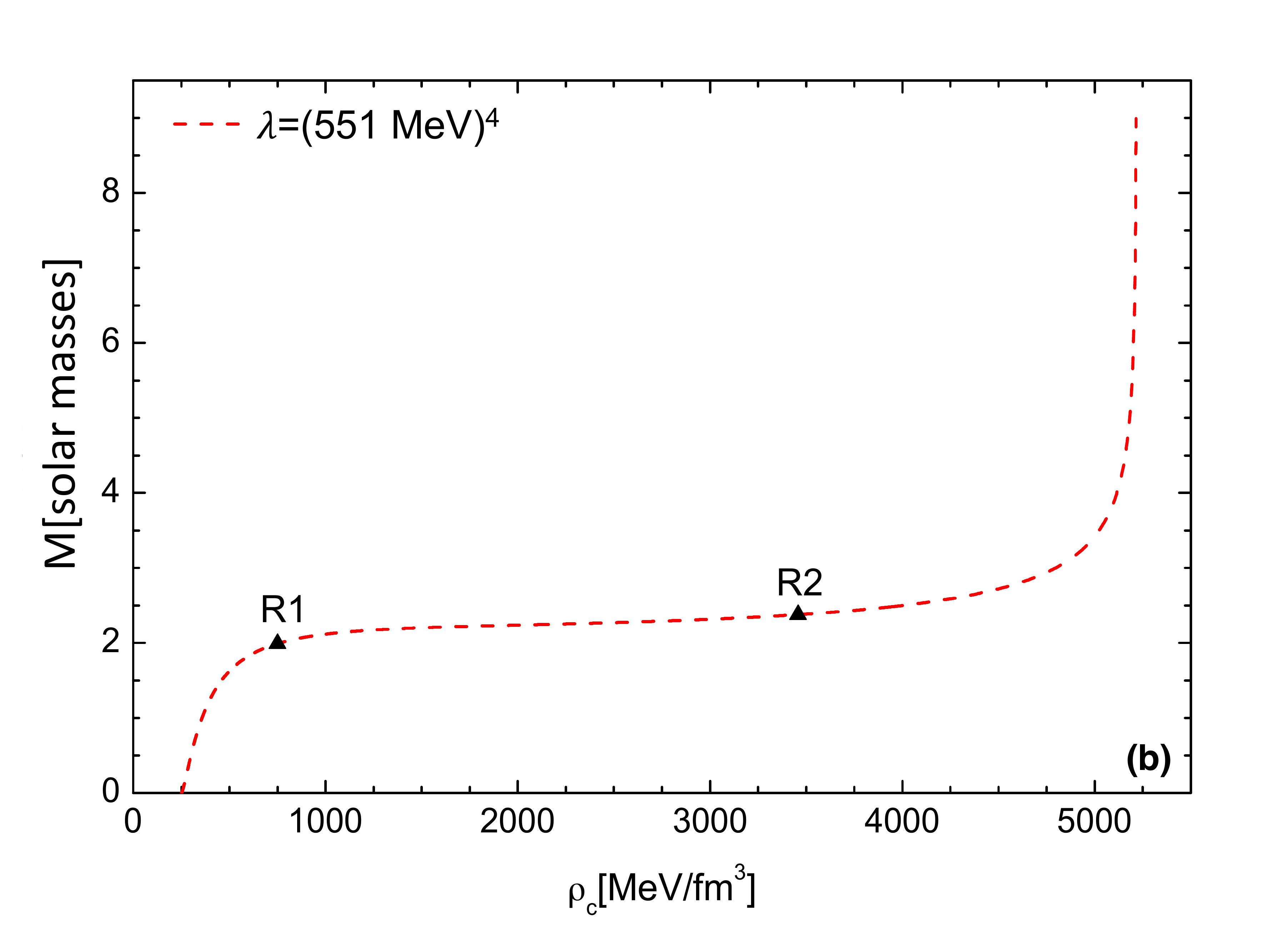}
\includegraphics[scale=0.24]  {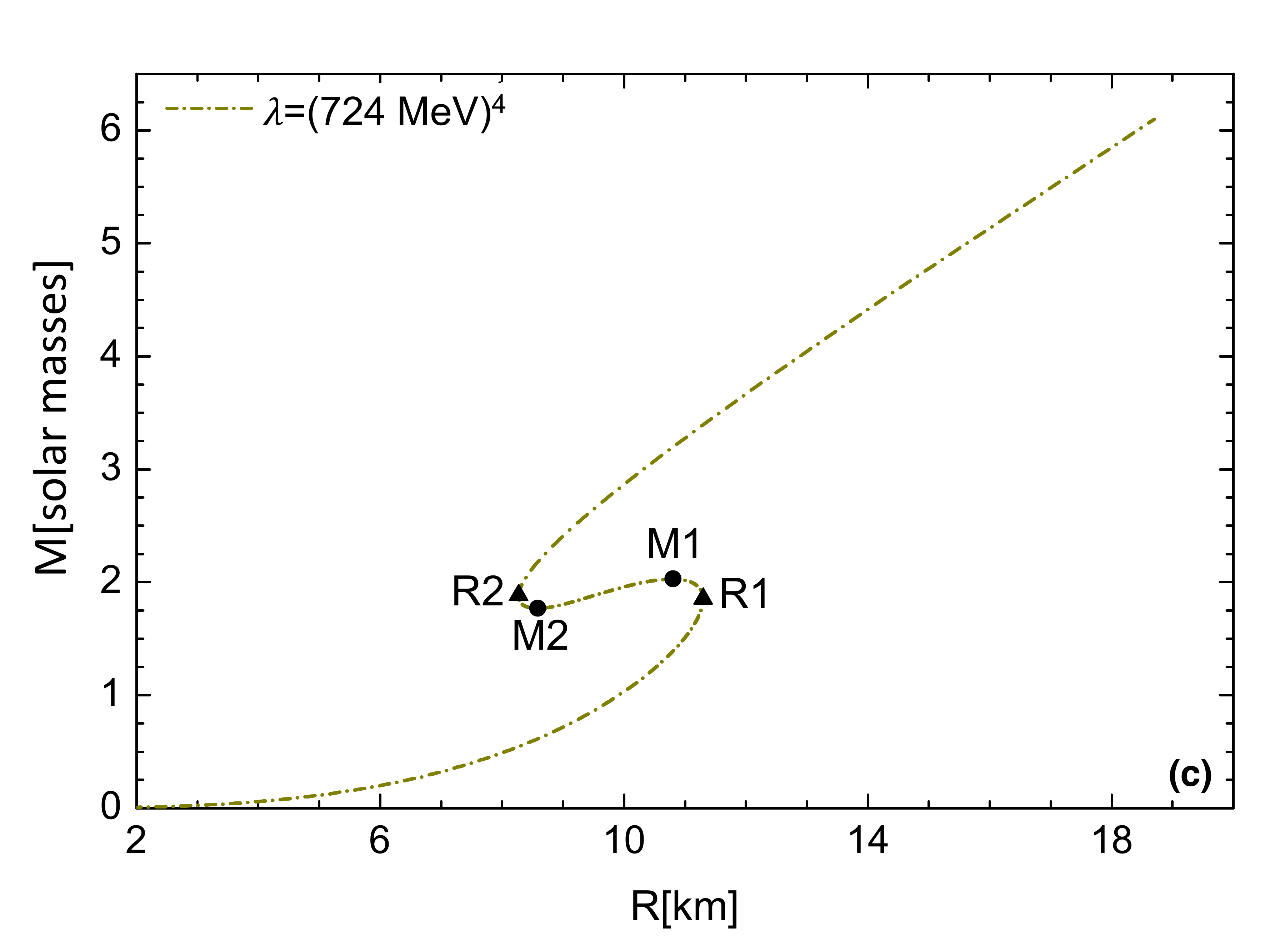}
\includegraphics[scale=0.24]  {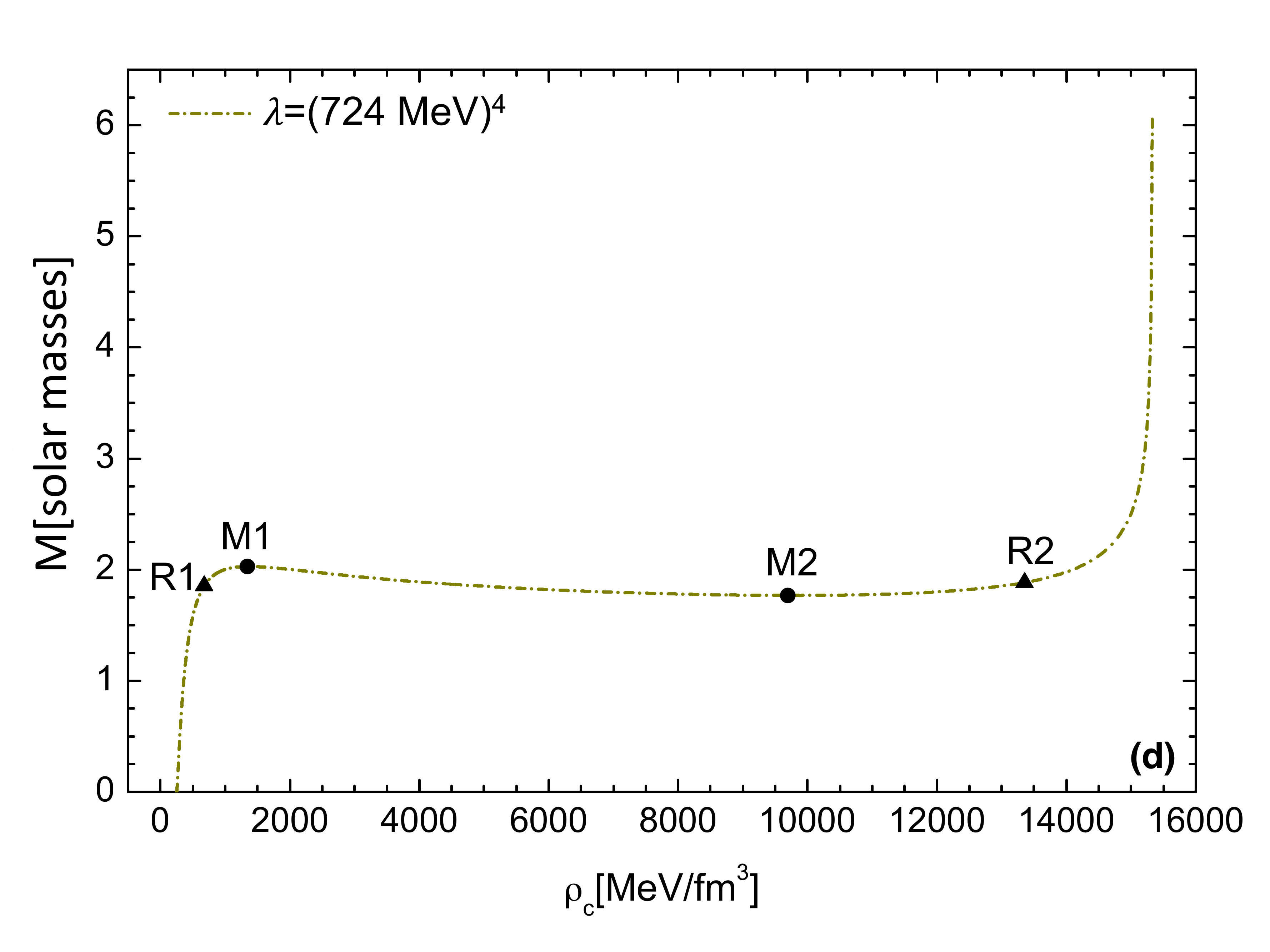}
\caption{Stability analysis of stellar configurations in braneworld models; for simplicity we present only the curves for strange quark stars. Figures (a) and (b) correspond to  a value of the brane tension, $\lambda=  (551 \, {\rm MeV})^4$, that results in no critical points. Figures (c) and (d) correspond to a different value of  the brane tension, $\lambda=  (724 \, {\rm MeV})^4$, that results in a local maximum and a local minimum in both the $M(R)$ and  $M(\rho_c)$ curves. For the analysis of the stellar stability based on these figures, see the text. }
\label{M_rhoc_stability}
\end{figure*}

In the previous sections we found that in braneworld models there is a new branch of stellar configurations that is not present within General Relativity. Since only stellar configurations  in stable equilibrium are acceptable from the astrophysical point of view,  we should check the stability of the previously obtained stellar models. 
A well known static criterion  that is widely used in the literature states that a necessary  condition for a model to be stable is that its mass $M$ increases with growing central density, i.e. 
\begin{equation}
\frac{d M}{d \rho_c} > 0 .
\end{equation}
The latter is a necessary but not sufficient condition. The opposite inequality $dM/d \rho_c < 0$ always implies instability of stellar models; i.e configurations lying on the segments with $dM/d \rho_c < 0$ are always unstable with respect to small perturbations.

In Figs. \ref{M_R} and \ref{M_rhoc}  there are two qualitatively different types of $M(R)$ and  $M(\rho_c)$ curves. One type  presents  one local maximum and one local minimum in both the  $M(R)$ and the $M(\rho_c)$ curves. The other one has no critical points. 
These two types are represented separately in Fig. \ref{M_rhoc_stability}, where we show  the $M(R)$ and  $M(\rho_c)$ curves for  strange quark stars for two different values of $\lambda$ (for simplicity we do not show hadronic stars because the stability analysis is completely equivalent, as we shall see below). 
For $\lambda=  (551 \, {\rm MeV})^4$ (upper panels) the stellar mass  is always an increasing function of the central density $\rho_c$  and the $M(R)$ curve has no local maxima or minima. Thus, the above necessary stability condition is always verified in this case.  
For  $\lambda=  (724 \, {\rm MeV})^4$  (lower panels)  the part of the $M(\rho_c)$ curve between the points $M1$ and $M2$ verifies $dM/d \rho_c < 0$, i.e. those configurations are unstable.  $M1$ indicates the local maximum  and $M2$ indicates the local minimum of the mass in both plots.  The  necessary stability condition is verified for the branch to the left of $M1$ and to the right of $M2$ in the $M(\rho_c)$ curve and in the corresponding branches of the $M(R)$ curve. 
Therefore, the branches that  approach asymptotically to the Schwarzschild limit  verify the necessary condition  $dM/d\rho_c > 0$ for any $\lambda$, but, as stated before, this is not a sufficient condition for stability.

A more detailed study of the stability of non-rotating spherically symmetric equilibrium models against small perturbations should be carried out through the analysis of their radial oscillations. However, this is left for future work because it is necessary to derive and solve the pulsation equations on the brane. Instead, we employ here a sufficient criterion which enables one to determine the precise number of unstable normal radial modes using the $M(R)$ curve \cite{htww,haensel}. According to such criterion, at each critical point of the $M(R)$ curve (local maxima or minima) one and only one normal radial mode changes its stability, from stable to unstable or vice versa. There are no changes of stability associated with radial pulsations at other points of the $M(R)$ curves. Moreover, one mode becomes unstable (stable) if and only if the $M (R)$ curve bends counterclockwise (clockwise) at the critical point.

In order to analyze the stellar stability using the above criterion,  we assume that the low mass branch (up to $\lesssim 1.5 - 2 M_{\odot}$) of the $M(R)$ curves is stable for all radial modes,  as it is in the general relativistic case.  
For the curves with two critical points,  the $M(R)$ curve bends counterclockwise at the local maximum and the fundamental oscillation mode becomes unstable. However, at the local minimum the fundamental mode becomes stable again because the curve bends clockwise there. Beyond the local minimum there are no more critical points and all the radial modes remain stable. 
In the case without critical points, the whole sequence remains stable for all radial modes provided that the low mass configurations are stable. Thus, we can conclude that  the branches that  approach asymptotically to the Schwarzschild limit are always stable under small radial perturbations. As a consequence, stellar configurations of arbitrarily large mass are allowed within braneworld models.

\section{Summary and Conclusions}
 \label{conclusions}

In this work we have studied the structure of compact stars in a Randall-Sundrum type II braneword model. To this end,  we employed the braneworld generalization of the stellar structure equations for a static fluid distribution with spherical symmetry. We considered that the spacetime outside the star is described by a Schwarzschild metric, i.e. the nonlocal pressure and energy density vanish outside the star, and assumed that the nonlocal pressure is zero in the stellar interior.  As a consequence, the interior must have nonvanishing nonlocal Weyl stresses ($\mathcal{U}^- \neq 0$).

In order to obtain an upper bound to the maximum mass of compact stars in the braneworld model, we integrated the stellar structure equations employing the causal limit EOS, which is obtained adopting the well established Baym, Pethick, and Sutherland  EOS  at densities below a fiducial density, and an EOS with the sound velocity equal to the speed of light above it. Assuming the causal limit EOS,  it can be shown that  the region above the causality limit depicted in Fig.~\ref{LC} is forbidden within General Relativity. However, the equilibrium solutions found in the braneworld model can violate the limit of causality for General Relativity and, for sufficiently large mass they approach asymptotically to the Schwarzschild limit $M = 2 R$.
Hence, there is a region in the $M-R$ plane that is forbidden in General Relativity for causal equations of state but that can be accessed within braneworld models; i.e.  the region between the straight lines shown in Fig.~\ref{LC}.

Then, we investigated the properties of hadronic and strange quark stars using two typical  EOSs that have been extensively employed in the literature: a nonlinear relativistic mean-field model for hadronic matter and  the MIT bag model for quark matter.  The main characteristics of the mass-radius relationship  found using the causal limit EOS are confirmed for both  hadronic and  strange quark stars.  For small masses ($\lesssim 1.5 - 2 M_{\odot}$),  the curves show the typical behavior found within the frame of General Relativity, i.e. very small mass hadronic stars have very large radii,  while strange stars follow roughly $M(R)  \propto R^3$.  Moreover,  the $M(R)$ curves for hadronic and quark stars  bend anticlockwise  at around $1.5 - 2 M_{\odot}$ as in the general relativistic case. However, the curves bend once more (clockwise) for larger masses and thereafter they increase roughly linearly and approach asymptotically to the Schwarzschild limit.   Again, two remarkable features are that  the $M(R)$ curves violate the general relativistic causal limit, and  that  static stellar configurations do not have a maximum mass as in the  general relativistic case, i.e. objects of any mass are possible in principle (see Fig. \ref{M_R}).  These large mass stars are supported against collapse by the nonlocal effects of the bulk on the brane.

Finally, we studied the stability under small perturbations of the stellar configurations in the braneworld. We used a static criterion which enables one to determine the precise number of unstable normal radial oscillation modes analyzing the bending of the mass-radius curves at the critical points (see Fig. \ref{M_rhoc_stability}). We assumed that the low mass branch (up to $\lesssim 1.5 - 2 M_{\odot}$) of the $M(R)$ curves is stable for all radial modes,  as it is in the general relativistic case. For the models without critical points, there are no changes of stability associated with radial oscillation modes, and therefore all configurations are stable. For the models with two critical points,  the $M(R)$ curve bends counterclockwise at the local maximum and the fundamental oscillation mode becomes unstable there. However, the fundamental mode regains stability at the  local minimum because the curve bends clockwise there. Beyond that minimum there are no more critical points and all the radial modes remain stable.

{In summary, within braneworld models we obtain the low mass branch of compact star configurations already known from general relativistic calculations, but we also find a new branch that approaches asymptotically to the Schwarzschild limit which is always stable under small radial perturbations. This new branch contains stellar configurations of arbitrarily large mass,  supported against collapse by the nonlocal effects of the bulk on the brane.} 

{It is worth emphasizing that black holes are still possible within the here studied braneworld models. Moreover, the stellar configurations that asymptotically approach to the Schwarzschild limit are expected to be stable under small perturbations, but not necessarily under large ones. Therefore, a very large mass braneworld compact star could collapse into a black hole if strongly perturbed in a catastrophic astrophysical event, e.g. in a binary stellar merging. }

{Finally, we remark that although a complete 5D analysis would be necessary to fully understand the properties of the new branch, these results serve as a proof of principle that traces of extra-dimensions might be found in astrophysics, specifically through the analysis of  masses and radii of compact objects. }

\begin{acknowledgments}

\noindent 
GL acknowledges the Brazilian agencies CNPq  and FAPESP for financial support.  JDVA thanks the Brazilian agencies CAPES and FAPESP (Project 13/26258-4).

\end{acknowledgments}

\appendix 
\section{Analytic solution for the nonlocal energy density ${\cal U}^{-}$ for a linear equation of state}
\label{appendixA}
%
For an arbitrary equation of state, the nonlocal energy density inside the star ${\cal U}^{-}(r)$ must be obtained through the numerical integration of Eq. \eqref{U-brane2} together with Eqs. \eqref{mass-brane} and \eqref{tov-brane}. However, as we show below, an analytic solution for ${\cal U}^{-}(r)$ can be obtained in the case of a linear EOS. First, we rewrite Eq. \eqref{U-brane2} in the form
\begin{equation}\label{eq1}
\frac{d{\cal U}^{-}(r)}{dr}+{\cal U}^{-}(r)g(r)= f(r),
\end{equation}
where
\begin{eqnarray}
&&g(r)=-\frac{4}{p(r)+\rho(r)}\frac{dp(r)}{dr},\\
&&f(r)=-2(4\pi G)^2(\rho(r)+p(r))\frac{d\rho(r)}{dr}.
\end{eqnarray}
Multiplying by the integrating factor $e^{\int g(r)dr}$ we have
\begin{equation}
\frac{d}{dr}\left({\cal U}^{-}(r)e^{\int g(r)dr}\right)= f(r)e^{\int g(r)dr},\label{eq3}
\end{equation}
which gives:
\begin{equation}\label{eq5}
{\cal U}^{-}(r)=e^{-\int g(r)dr}\left[\int f(r)e^{\int g(r)dr}dr+C_1\right],
\end{equation}
where $C_1$ is an integration constant. 

Now, we consider a linear EOS of the form  $\rho=p/c^2_{s}+b$, where $c^2_s$  and $b$ are constants ($c_s$ is the speed of sound). Notice that this EOS includes as special cases the \textit{causal} EOS $\rho=p$,  the \textit{ultra-relativistic} EOS $\rho=3 p$, and the \textit{MIT bag model} EOS for massless quarks $\rho = 3 p + 4 B$. For such EOS,  Eq. \eqref{eq5} reads
\begin{eqnarray}\label{eq_int}
{\cal U}^{-}(r) &=&\left[ (1 + c^2_s) p(r) + b c^2_s\right]^{\frac{4c^2_s}{(1+c^2_s)}}\left[-\frac{2(4\pi G)^2}{c^4_s}\times\right.\nonumber\\
&&\left.\int\left[ (1+c^2_s)  p(r) + b c^2_s\right]^{\frac{1-3\,c^2_s}{1+c^2_s}} dp + C_1\right],
\end{eqnarray}
with $k_1$ being an integration constant that comes from the integral $\int g(r)dr$. 
Integrating Eq. \eqref{eq_int} we find
\begin{equation}\label{eq9}
{\cal U}^{-}(r)=\left\{\begin{split}
&[2p(r)+b]^2\left[-(4\pi G)^2\ln(2p(r)+b)\right.\\
&\left.+k_2\right], \quad c_s=1;\\
&\left[(1+c^2_s)p(r) + b c^2_s \right]^{\frac{4\,c^2_s}{(1+c^2_s)}}\left[-\frac{(4\pi G)^2}{(1-c^2_s)}\times\right.\\
&\left.\left[ (1+c^2_s) p(r)+ b c^2_s\right]^{\frac{2(1-c^2_s)}{(1+c^2_s)}} + k_2\right], \quad c_s<1;
\end{split} \right. 
\end{equation}
where $k_2=C_2+k_1\,C_1$, being $C_2$ another integration constant.

To determine $k_2$ we use the boundary condition:
\begin{equation}\label{eq10}
(4\pi G)^2\rho^2(R)+{\cal U}^{-}(R)=0.
\end{equation}
The value ${\cal U}^{-}(R)$ is obtained evaluating Eq. \eqref{eq9} at the surface of the star, and from the EOS we obtain $\rho(R) = b$ because $p(R) =0$. Therefore, Eq. \eqref{eq10} reads:
\begin{equation}\label{eq12}
k_2=\left\{\begin{split}
&(4\pi G)^2\left[\ln(b)-1\right], \quad c_s=1;\\
& (4\pi\,G)^{2} b \frac{ (b c_s^2)^{\frac{(1- 3 c^2_s)}{(1+c^2_s)}}}{(1-c^2_s)}, \quad c_s<1.
\end{split} \right. 
\end{equation}

Replacing $k_2$ from  Eq. \eqref{eq12}  into Eq. \eqref{eq9}  we find
\begin{equation}
{\cal U}^{-}(r)=\left \{\begin{split}
&-(4\pi G)^2[2p(r)+b]^2\left[\ln(2p(r)+b)\right.\\
&\left.-\ln(b)+1\right], \quad c_s=1;\\
&-\frac{(4\pi c_s G)^2}{(1-c^2_s)}\left[\frac{1}{c^2_s}\left[\frac{(1+c^2_s)}{c^2_s}p(r)+b\right]^{2}\right.\\
&\left.-b^{\frac{2(1-c^2_s)}{(1+c^2_s)}}\left[\frac{(1+c^2_s)}{c^2_s}p(r)+b\right]^{\frac{4c^2_s}{(1+c^2_s)}}\right],\; c_s<1;
\end{split} \right. 
\end{equation}
\noindent which gives ${\cal U}^{-}(r)$ as a function of $p(r)$, where $p(r)$ is to be obtained from the integration of Eqs. \eqref{mass-brane} and \eqref{tov-brane}.
An equivalent expression of  ${\cal U}^{-}(r)$ as a function of $\rho(r)$ can be obtained using the EOS:

\begin{equation}
{\cal U}^{-}(r)=\left\{\begin{split}
&-(4\pi G)^2[2\rho(r)-b]^2\left[\ln(2\rho(r)-b)\right.\\
&\left.-\ln(b)+1\right], \quad c_s=1;\\
&-\frac{(4\pi c_sG)^2}{(1-c^2_s)}\left[c^2_s\left[\frac{(1+c^2_s)}{c^2_s}\rho(r)-b\right]^{2}\right.\\
&\left.-b^{\frac{2(1-c^2_s)}{(1+c^2_s)}}\left[(1+c^2_s)\rho(r)-bc^2_s\right]^{\frac{4c^2_s}{(1+c^2_s)}}\right], c_s<1.
\end{split} \right. 
\end{equation}
In particular, we can evaluate the latter expression at the center of the star in order to obtain a relationship between the central nonlocal energy density ${\cal U}^{-}_{c} \equiv {\cal U}^{-}(0)$ and the central mass-energy density $\rho_c \equiv \rho(0)$
\begin{equation}\label{ENL_c}
{\cal U}^{-}_c=\left\{\begin{split}
&-(4\pi G)^2[2\rho_c-b]^2\left[\ln(2\rho_c-b)\right.\\
&\left.-\ln(b)+1\right], \quad c_s=1;\\
&-\frac{(4\pi c_s\,G)^2}{(1-c^2_s)}\left[c^2_s\left[\frac{(1+c^2_s)}{c^2_s}\rho_c-b\right]^{2}\right.\\
&\left.-b^{\frac{2(1-c^2_s)}{(1+c^2_s)}}\left[(1+c^2_s)\rho_c-bc^2_s\right]^{\frac{4c^2_s}{(1+c^2_s)}}\right],\; c_s<1.
\end{split} \right. 
\end{equation}

\end{document}